\documentclass[notitlepage,reprint,showpacs,nobalancelastpage,twocolumn,aps,prb]{revtex4-2}
\RequirePackage[T1]{fontenc}
\RequirePackage{times}
\usepackage{amsfonts,amssymb,amsmath}
\usepackage{bm}
\usepackage{my_symbols}
\usepackage[italicdiff]{physics}
\usepackage{xcolor}
\usepackage{lipsum}
\usepackage{CJK}
\usepackage[english]{babel}
\usepackage{graphicx}
\usepackage{subfigure}
\usepackage{siunitx}
\usepackage{float}
\usepackage[normalem]{ulem}
{\par}
\newcounter{mycomment}

\newcommand\rmv{\bgroup\markoverwith {\textcolor{red}{\rule[0.5ex]{2pt}{0.4pt}}}\ULon}

\begin{document}
\begin{CJK*}{UTF8}{gbsn}
\title{Curvilinear manipulation of polarized spin wave}
\author{Hongbin Wu (武宏斌)}
\author{Jin Lan (兰金)}
\email[Corresponding author:~]{lanjin@tju.edu.cn}
\affiliation{Center for Joint Quantum Studies and Department of Physics, School of Science, Tianjin University, 92 Weijin Road, Tianjin 300072, China}

\begin{abstract}
Polarization, the precession direction with respect to the background magnetization, is an intrinsic degree of freedom of spin wave.
Introducing symmetry breaking mechanisms lies in the heart of lifting the degeneracy between polarized spin wave modes, and is essential in constructing polarization-based magnonic devices.
Here we show that polarized spin waves can be naturally harnessed in a curved antiferromagnetic wire via tuning its curvature and torsion.
Specifically, we investigate evolution of polarized spin wave in a spin wave rotator and a spin wave interferometer based on magnetic circular helices, and correlate these curvilinear effects to the Berry phase accumulated along wires.
\end{abstract}
\maketitle
\end{CJK*}

\section{Introduction}
Spin wave, the collective precession of ordered magnetizations, is one of the basic  excitations in magnetic systems.
As an alternative angular momentum carrier beside spin-polarized electron \cite{kajiwara_Transmission_2010,cornelissen_Longdistance_2015}, spin wave is of crucial importance for both fundamental physics and industrial applications.
And since the propagation of spin wave does not involve physical motion of underlying electrons, magnonics as a field devoted to spin wave manipulation has attracted remarkable interests in recent years \cite{kruglyak_Magnonics_2010,chumak_Magnon_2015,lan_SpinWave_2015,han_Mutual_2019,yu_Magnetic_2020,yu_Magnetic_2021,barman_2021_2021}.

Polarization, denoting the oscillation direction, is an intrinsic property for all waves, including electromagnetic, acoustic as well as spin waves \cite{keffer_Theory_1952,cheng_Antiferromagnetic_2016,lan_Antiferromagnetic_2017}.
Theoretically, polarization has emerged as a key ingredient in spin wave field-effect transistor \cite{cheng_Antiferromagnetic_2016}, spin wave polarizer and retarder \cite{lan_Antiferromagnetic_2017}, spin wave double refraction \cite{lan_Geometric_2021},  magnonic spin Nernst effect \cite{cheng_Spin_2016,zyuzin_Magnon_2016}, as well as using spin wave to drive magnetic texture motion \cite{tveten_Antiferromagnetic_2014,qaiumzadeh_Controlling_2018,yu_Polarizationselective_2018,daniels_Topological_2019,oh_Bidirectional_2019a}.
In the meantime, recent rapid technical improvements enable the detection of polarized spin wave in various experiments, including antiferromagnetic spin pumping \cite{cheng_Spin_2014,vaidya_Subterahertz_2020,li_Spin_2020a}, inelastic neutron scattering \cite{gitgeatpong_Nonreciprocal_2017a,nambu_Observation_2020}, antiferromagnetic resonance \cite{macneill_Gigahertz_2019} and  spin signal transmission \cite{han_Birefringencelike_2020}.

Multiple approaches have been proposed to harness the polarized spin wave, e.g. applying external magnetic field \cite{keffer_Theory_1952}, introducing the Dzyaloshinskii-Moriya interaction \cite{cheng_Antiferromagnetic_2016}, depositing magnetic textures \cite{lan_Antiferromagnetic_2017,daniels_Topological_2019,lan_Geometric_2021}, and streaming electrical current \cite{proskurin_SpinWave_2017}.
However, the applicability and efficiencies of these approaches are frequently limited by the magnetic properties and configurations available in realistic materials.
A convenient scheme to lift the restrictions is to use curvilinearity embedded in bent wires or curved surfaces, which are ready to prepare in state of art experiments \cite{balhorn_SpinWave_2010,smith_Magnetic_2011,dacol_Observation_2014,kim_Synchronous_2014,streubel_Magnetism_2016,volkov_Experimental_2019}.
Despite of this unique yet powerful scheme in modifying the magnetic properties \cite{yan_Fast_2011,gaididei_Curvature_2014,sheka_Curvature_2015,otalora_CurvatureInduced_2016},  possibilities of using curvilinear effects to harness polarized spin wave are not fully explored.

In this work, we investigate the dynamics of polarized spin wave along a curved antiferromagnetic wire, using the language of curvature and torsion.
We show that for a magnetic circular helix shifts the phase between left/right circular spin wave modes, due to the emergent Dzyaloshinskii-Moriya interaction.
We further propose that a spin wave rotation based on a single cicular helix, and a spin wave interferometer based two oppositely wound circular helices.
Using curvilinear effects for polarization manipulation, offers new paradigm in constructing magnonic logic devices.

The rest of paper is organizeds as follows.
The magnetic dynamics in curved wire is formulated by establishing the connections between curvilinear effect to an emergent Dzyaloshinskii-Moriya interaction in Sec. II.
The phase shift of spin wave experienced in circular helices, the functionalities of  spin wave rotator and spin wave interferometer are studied in Sec. III.
Discussions about the connection between torsion and Berry phase, as well as a short conclusion,   are  given in  Sec. IV.

\section{Basic model}
\subsection{Magnetic dynamics in a curved antiferromagnetic wire}
\begin{figure}[bt]
\centering
{\includegraphics[width=0.45 \textwidth]{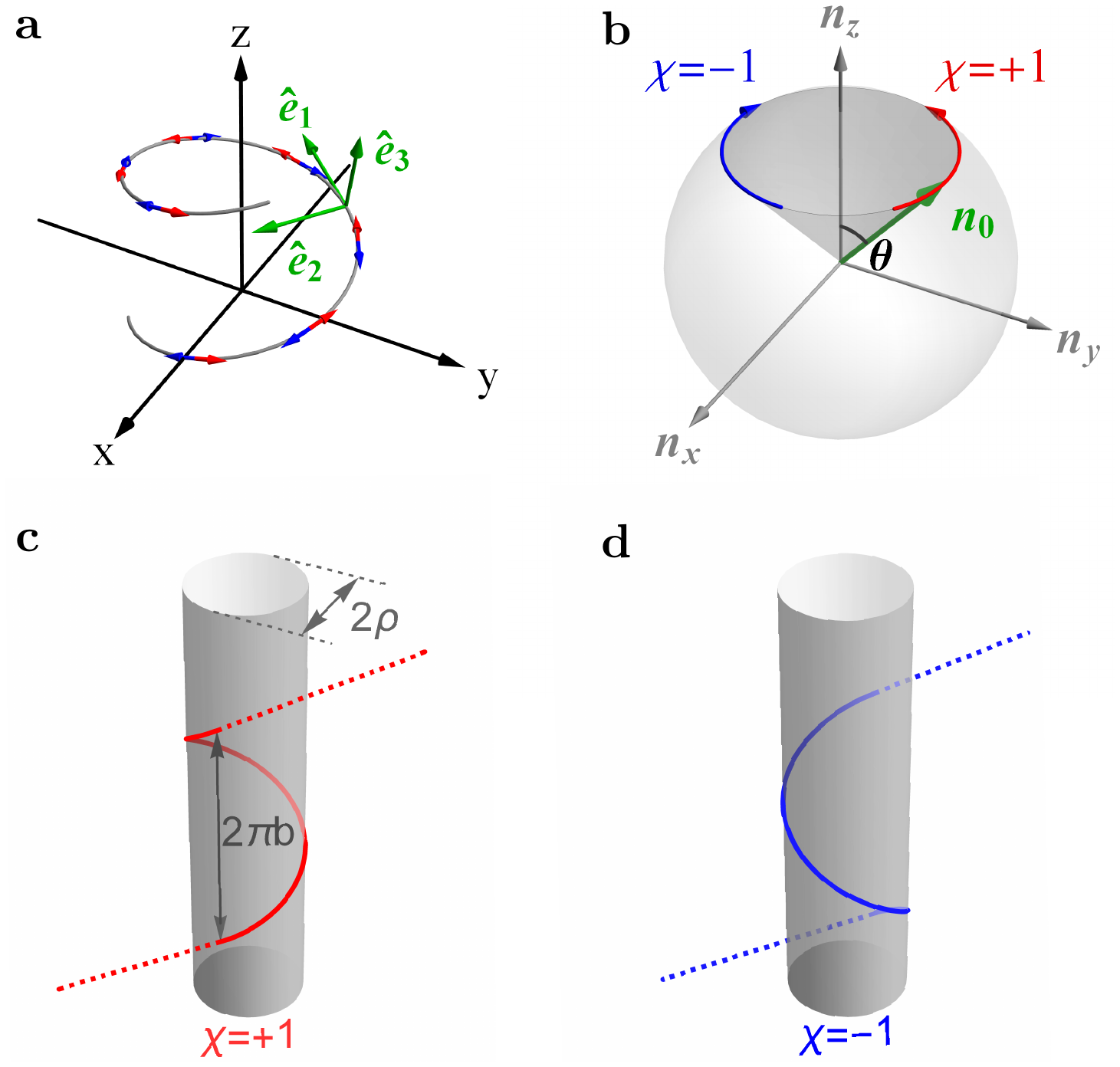}}

\caption{{\bf Schematics of curved magnetic wires.}
(a) Magnetizations in a curved magnetic wire. The gray line is the curved magnetic wire, the red/blue arrows are the magnetizations in two sublattices, and three green arrows are the curvilinear bases.
(b) Magnetizations in a magnetic Bloch sphere.
The anticlockwise/clockwise  winding with helicity $\chi=\pm 1$ are shown in red/blue color.
(c)(d) Space curves of left/right wound circular helices.   Dashed lines are straight wires in smooth connection with the circular helix.
\label{fig:sch_curved_afm}
}
\end{figure}

Consider a curved antiferromagnetic wire embedded in $3$D space as shown in Fig. \ref{fig:sch_curved_afm}(a), which is described by the parameterized position $\br(s)$ with the parameter $s$ being the arc length of the wire.
To describe the dynamics in a curved wire,  it is convenient to introduce local curvilinear bases in the Frenet-Serret frame \cite{gaididei_Curvature_2014,sheka_Curvature_2015},
\begin{align}
\label{eqn:curvi_bases}
       \hbe_1 = \frac{\br'}{| \br '|}, \quad \hbe_2 = \frac{ \hbe_1'}{| \hbe_1' |}, \quad \hbe_3= \hbe_1 \times \hbe_2,
\end{align}
where $\br'\equiv d\br/ds$ denotes the derivative with respect to the arc length $s$,  $\hbe_{1/2/3}$ are the tangent, normal and binormal vectors of the curve $\br(s)$ respectively.

We denote the magnetization directions of two sublattices in the curved antiferromagnetic wire by unit vectors $\mb_{1/2}$, as depicted by red/blue arrows in Fig. \ref{fig:sch_curved_afm}(a).
The magnetizations are alternatively described by the staggered N\'{e}el order magnetization $\bn=(\mb_1-\mb_2)/(|\mb_1-\mb_2|)$ and the net magnetization as $\mb=\mb_1+\mb_2$.
Under the orthogonal constraint $\bn\cdot \mb=0$, the dynamics of the staggered order $\bn$ is then governed by \cite{haldane_Nonlinear_1983,tveten_Staggered_2013,kim_Propulsion_2014,yu_Polarizationselective_2018}
\begin{align}
\label{eqn:eom_afm}
          \ddot{\bn} \times \bn = -\gamma  J  \bn\times \qty(\gamma \bh  - \alpha \dot{\bn}),
\end{align}
and $\mb=(\bn\times \dot{\bn})/J$ is a slave quantity.
Here $\ddot{\bn}\equiv \partial_t^2 \bn$, $\dot{\bn}\equiv \partial_t \bn$,  $\gamma$ is the gyromagnetic constant, $\alpha$ is the Gilbert damping constant, and $\bh =-\delta U/\delta \bn$ is the effective field acting on staggered magnetization $\bn$.
The magnetic energy of interest in this work is $U=\int u  \, d\br $ with the energy density described by
\begin{align}
\label{eqn:mag_en}
        u(\bn,\mb)=&u_\ssf{A}+u_\ssf{K}+u_m \nonumber\\
         =& \frac{1}{2}  [A (\nabla \bn)^2- K (\bn\cdot \hbe_1)^2  +  \frac{J}{2} \mb^2 ],
\end{align}
where  $A$ and $J$ are the intra/inter-sublattice exchange coupling constants, $K>0$ is the easy-axis anisotropy along the tangential direction $\hbe_1$ of the wire.
In this work, the long range dipolar interaction is neglected due to the interweaving sublattices with almost opposite magnetizations in the antiferromagnetic environment.

\subsection{Emergent Dzyaloshinskii-Moriya interaction and anisotropy}
Continuous variations of the curvilinear bases $\hbe_{1/2/3}$ in \Eq{eqn:curvi_bases} render the physical properties of a curved wire distinct from a straight wire.
The differential properties of a curved wire is characterized by the  Frenet-Serret  formula $  \hbe_1 ' =\kappa \hbe_2$, $  \hbe_2'=-\kappa \hbe_1+\tau \hbe_3$, and $ \hbe_3'=-\tau \hbe_2 $,  where $\kappa$ and $\tau$ are the curvature and torsion of the curved wire \cite{patrikalakis_Shape_2002}, respectively.
The curvature $\kappa$ measures the bending of the wire within a plane, while the torsion $\tau$ measures the twist of the wire out of a specific plane.
In vector form, the Frenet-Serret formula reads
\begin{align}
	\label{eqn:fs_formula}
	 \hbe'_\beta  =\bd \times \hbe_\beta,
\end{align}
where  $\bd =(\tau \hbe_1+\kappa \hbe_3)$ is the curvilinear vector induced by the spatial variation of the curvilinear basis $\qty{\hbe_\beta}, \beta={1, 2, 3}$.
The chiral nature of  \Eq{eqn:fs_formula} implies that the magnetic system in a curved wire  resides  in a local rotating frame with respect to vector $\bd$ \cite{goldstein_CLASSICAL_2000}.

Invoking the Frenet-Serret formula in \Eq{eqn:fs_formula}, the derivative of the staggered magnetization $\bn=\sum_{\beta} n_\beta \hbe_\beta$ then takes following co-derivative form
\begin{align}
\label{eqn:chiral_der}
      \bn'\equiv \dv{\bn}{s} = \qty(\partial_s  + \bd \times) \bn,
\end{align}
where  $\partial_s \bn  \equiv \sum_\beta (\partial_s n_\beta) \hbe_\beta$ denotes the partial derivative solely acting on the magnetization component $n_\beta$.
Including the co-derivative in \Eq{eqn:chiral_der},  the exchange coupling energy $u_\ssf{A}$ in \Eq{eqn:mag_en} then divides into following $3$ parts \cite{pylypovskyi_Curvilinear_2020},
\begin{align}
\label{eqn:emerg_DMI}
        u_\ssf{A}  =&          \tilde{u}_\ssf{A} + \tilde{u}_\ssf{D} +\tilde{u}_\ssf{K} \nonumber \\
        =&   \frac{1}{2} \qty[ A(\partial_s \bn)^2 + 2 A  \bd\cdot (\bn\times  \partial_s \bn) - A (\bd\cdot \bn)^2],
\end{align}
where an effective Dzyaloshinskii-Moriya interaction (DMI) in vector $\bD=2A\bd$, as well as an effective easy-axis anisotropy of strength $A$ with respect to vector $\bd$,   arise besides the normal exchange coupling in terms of $\partial_s\bn$ \cite{gaididei_Curvature_2014,sheka_Curvature_2015,streubel_Magnetism_2016}.

In \Eq{eqn:emerg_DMI}, the Dzyaloshinskii-Moriya (DM) vector is explicitly written as
\begin{align}
\bD= 2A\tau\hbe_1+2A\kappa \hbe_3,
\end{align}
indicating that the torsion $\tau$ and curvature $\kappa$ determine the DM strength with respect to the tangential/binormal directions $\hbe_{1/3}$, respectively.
The strength of emergent DMI in \Eq{eqn:emerg_DMI} is proportional to the exchange coupling constant $A$, similar to the DMI caused by the spin-orbit coupling in heavy-metal materials or the hetero-structure interface \cite{kim_Chirality_2013}.
In addition, the DMI caused by curvilinear effects here  also shares much similarity with the geodesic effects due to geometric spacetime distortion in general relativity \cite{hill_Chiral_2021}.

On the other hand, from \Eq{eqn:mag_en} and \Eq{eqn:emerg_DMI}, the overall energy density of magnetic anisotropy is described by
\begin{align}
\label{eqn:ani}
u_\ssf{K}+\tilde{u}_\ssf{K}=-\frac{K}{2} n_1^2 -\frac{A}{2} (\tau n_1+\kappa n_3)^2,
\end{align}
where the torsion $\tau$ leads to an enhanced easy-axis anisotropy $K\to K +A\tau^2$, while the curvature $\kappa$ causes a small tilting of the  magnetic easy axis toward the binomial direction $\hbe_3$.

In experimental practice, the wire is supposed to bend only slowly in comparison the characteristic length of the magnetic system: $\kappa,\tau \ll 1/W$ with $W\equiv \sqrt{A/K}$, hence the relation $A\kappa^2\ll K$ gives rise to negligible tilting of the magnetic easy axis.
In the meantime, the anisotropy also dominates the emergent DMI, therefore homogeneous magnetizations along the magnetic easy-axis $\bn_0=\hbe_1$ maintain to be the ground state in the curved wire.
Inhomogeneous magnetic textures such as domain wall may also arise upon curved magnetic wire, but are beyond the scope of this work and thus omitted.

\subsection{Dynamics of spin wave in a curved antiferromagnetic wire}
When time evolution is involved, the total staggered magnetization $\bn(t)$  naturally separates into the static background $\bn_0$ and the fast precessing spin wave $\delta \bn$: $\bn=\bn_0+\delta{\bn}$.
Due to unity constraint $|\bn|=1$, the spin wave is orthogonal to the background magnetization with $\bn_0\cdot \delta{\bn}=0$ satisfied everywhere.
Upon the homogenous magnetization $\bn_0=\hbe_1$ along the tangential direction, the spin wave is written as $\delta{\bn}=n_2 \hbe_2+ n_3 \hbe_3$, i.e. spin wave oscillates in the normal and binormal directions of the wire.
In an equivalent form, the spin wave is denoted by complex field  $\psi_\sigma =n_2 - i \sigma n_3$, with the chirality $\sigma=\mp  1$ the left/right circularly polarized spin waves, respectively.
Adopting the small amplitude limit $|n_{2/3}|\ll 1$, the spin wave dynamics in the curved wire is recast from \Eq{eqn:eom_afm} to
\begin{align}
   \label{eqn:sw_eom}
  -\ddot{\psi}_{\sigma}=\gamma^2 J\qty[ A(i\partial_s+\sigma\tau)^{2}+K ] \psi_{\sigma},
\end{align}
where the degeneracy of two circular polarizations are lifted by the torsion $\tau$.
The curvature $\kappa$ does not participate into \Eq{eqn:sw_eom}, since the DMI orthogonal to the static magnetization direction does not affect the spin wave dynamics in linear regime \cite{yu_Magnetic_2016}.

The spin wave dispersion corresponding to \Eq{eqn:sw_eom} is
\begin{align}
\label{eqn:sw_disp}
\omega_\sigma=\gamma \sqrt{J\qty[ A (  k_\sigma-\sigma \tau)^2 +K ]},
\end{align}
where $\omega_\sigma$ and $k_\sigma$ with $\sigma=\mp 1$ are the spin wave frequency and wavevector for left/right circular modes, respectively.
In comparison, for a straight wire with vanishing torsion $\tau=0$, the dispersion reduces to $\omega_0=\gamma \sqrt{J\qty ( A   k_0^2 +K )}$.
Therefore, in reference to the straight wire with fixed spin wave frequency $\omega$, the wavevector of spin wave in a curved wire is  shifted by \cite{cheng_Antiferromagnetic_2016}
\begin{align}
\Delta k\equiv k_\sigma-k_0 =\sigma \tau,  \label{eqn:delta_k}
\end{align}
 i.e., the wavevector of the left/right circular modes with $\sigma=\mp 1$ are shifted in opposite directions by the same amount of $|\tau|$.
It is noteworthy that the wavevector shift $\Delta k$ is independent of  the spin wave frequency $\omega$ and the detailed magnetic parameters $A$, $K$ and $J$ in \Eq{eqn:mag_en}.

\section{Polarized spin wave along circular helices}

\subsection{Basic properties of a circular helix}
The simplest wire with finite curvature and torsion is a circular helix \cite{ONEILL20061}, which is virtually wound upon a cylinder, as schematically shown in Fig. \ref{fig:sch_curved_afm}(c)(d).
The  $3$D  space curve of a circular helix is parameterized by
\begin{align}
\label{eqn:helix_para}
 \br(s) = \qty( \rho \cos(\frac{s}{r}), \chi \rho \sin(\frac{s}{r}),  \frac{b s}{r} ),
\end{align}
where $\rho$ is the radius,  $b$ is the reduced  pitch, $r=\sqrt{\rho^2+b^2}$ is the parametric radius, and $\chi=\mp 1$ is the helicity for left/right-handed helices.
The tangent vector of a circular helix is described by $\hbe_1= \qty( -\rho\sin(s/r),\chi\rho\cos(s/r), b)/r$, which has a periodicity of $2\pi r$, and all tangent vectors form a constant polar angle $\theta=\arccos(\hbe_1\cdot\hbz)=\arccos(b/r)$ with respect to the $\hbz$ direction.
Therefore, a collection of magnetizations lying in tangent vectors of circular helix forms a cone in a magnetic Bloch sphere with apex semiangle $\theta$, as shown in \ref{fig:sch_curved_afm}(b).

From \Eq{eqn:helix_para}, the curvature and torsion of a helix are constants with
\begin{align}
\label{eqn:kappa_tau}
\kappa=\frac{\rho}{r^2}, \quad \tau=\frac{\chi b}{r^2}
\end{align}
where the curvature $\kappa$ is controlled the radius $\rho$, while the torsion $\tau$ is related to the reduced pitch $b$ as well as the helicity $\chi$.
The curvature and torsion are intimately connected with the polar angle $\theta$ via the relation
$\kappa r = \sin \theta$ and  $\tau r=\chi \cos\theta$,
hence the curvature $\kappa$ (torsion $\tau$) increases/decreases  as the polar angle $\theta$ increases.

According to  \Eq{eqn:delta_k} and \Eq{eqn:kappa_tau}, when a circular spin wave with chirality $\sigma=\pm 1$ travels along the circular helix with helicity $\chi=\pm 1$, the phase shift accumulated in one coil of length $2\pi r$  is then
\begin{align}
\label{eqn:phase_diff}
\Delta \phi  = \int_0^{2\pi r} \Delta k \,ds  = 2\pi r \sigma \tau=2\pi \sigma \chi \cos\theta,
\end{align}
which is simultaneously controlled by both spin wave chirality $\sigma$  and the winding helicity $\chi$.
By flipping either the spin wave chirality  $\sigma$ or the winding helicity $\chi$, the phase shift $\Delta\phi$ is flipped;
or alternatively speaking, the phase difference between two opposite chirality or helicity is $2|\Delta\phi|=4\pi\cos\theta$.
With this additional phase $\Delta\phi$, we proceed to investigate the evolution of polarized spin wave advancing along magnetic circular helices.

\subsection{Spin wave rotator with a single circular helix}
\begin{figure}[bt]
\centering
	%\frame{\includegraphics[width=0.23 \textwidth,trim=140 80 90 100,clip]{figs/curved_AFM.pdf}}
\includegraphics[width=0.5 \textwidth]{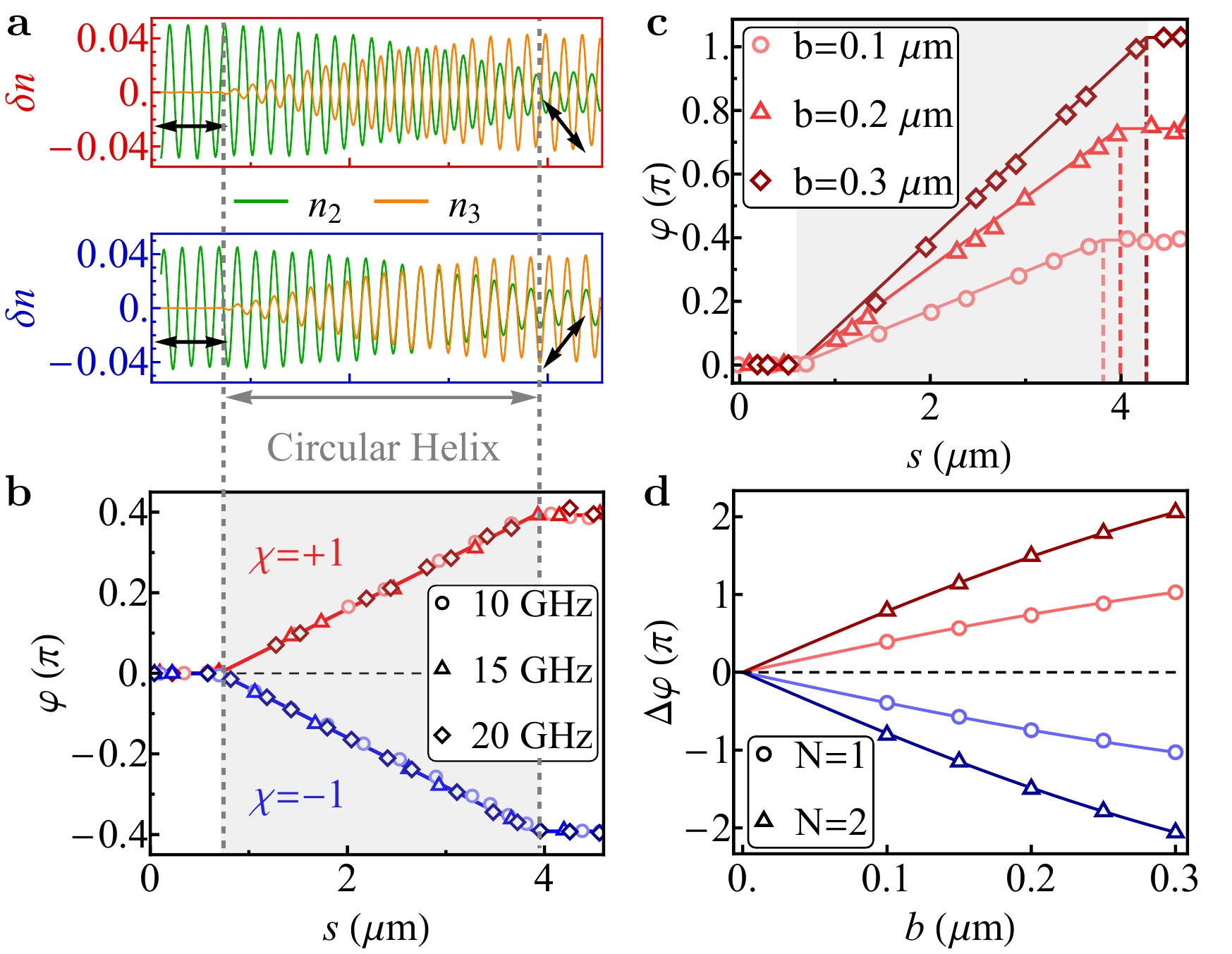}
\caption{
{\bf Micromagnetic simulations of spin wave rotation based on a magnetic circular helix.}
(a) Spatial profiles of spin wave across a magnetic helix with $\chi=\pm 1$ in upper/lower panels.
Two spin wave components $n_{2/3}$ are in plotted green/orange lines, and the arrows denote the polarization direction in two sides.
(b) Spatial evolutions of polarization angle $\varphi$ as function of arc length $s$ for three different frequencies $f=10, 15, 20~\si{GHz}$.
(c) Spatial profiles of polarization angle $\varphi$ for different reduced pitch $b$.
(d) The overall rotation angle $\Delta\varphi$ as function of reduced pitch $b$ for $1$ ($2$) coils of magnetic helices.
For (b)(c)(d), the dots are extracted from micromagnetic simulations,  the lines are theoretical calculations based on  \Eq{eqn:delta_varphi}, and the red/blue lines are for magnetic helix with helicity $\chi=\pm 1$.
The default size of the circular helix is with the radius $\rho=0.5~\si{\mu m}$ and the reduced pitch $b=0.1~\si{\mu m}$.
\label{fig:helix_sim}
}
\end{figure}
Consider a linearly polarized spin wave, which consists of equal components of both circular polarizations, propagates along a magnetic circular helix.
According to \Eq{eqn:phase_diff}, the left/right circular components experience opposite phase shifts $\pm |\Delta\phi|$, leading to a rotation of the polarization direction for the linear spin wave \cite{datta_Electronic_1990,cheng_Antiferromagnetic_2016,proskurin_SpinWave_2017,yu_Magnetic_2020}, i.e. spin wave Faraday effect arises in a magnetic circular helix.
The rotation angle of the linear spin wave, which takes half of the phase difference between its left and right circular components, is then described by
\begin{align}
\label{eqn:delta_varphi}
       \Delta \varphi=  2\pi \chi \cos\theta,
\end{align}
where the rotation angle $\Delta\varphi$ flips its sign for oppositely wound helices with $\chi=\pm 1$.
The rotation angle is directly proportional to the polar angle $\theta$ of the helix, highlighting the critical role of torsion (or the emergent DMI) in spin wave rotation.

To study the spin wave dyanmics with more details, we turn to micromagnetic simulations using Comsol Multiphysics \cite{COMSOL}, where the LLG equation is transformed to a weak form.
Specifically, a synthentic antiferromagnet wire is used in simulations, which consists of two antiferromagnetically  coupled   ferromagnetic yttrium iron garnet $\mathrm{Y_3Fe_5O_{12}}$ (YIG) wires \cite{yu_Polarizationselective_2018,lan_Antiferromagnetic_2017,lan_Geometric_2021}.
The magnetic parameters adopted in the simulations are: the intra-layer exchange coupling constant $A=3.28\times 10^{-11} \mathrm{A}\cdot\mathrm{m}$,  the inter-layer coupling constant $J=5\times 10^5 ~ \mathrm{A}/\mathrm{m}$, the anisotropy $K= 3.88 \times 10^{4}~ \mathrm{A}/\mathrm{m}$, and the damping constant $\alpha=1\times 10^{-4}$.

The evolutions of a linearly polarized spin wave across a circular helix  between two straight wires, as extracted from micromagnetic simulations,  are plotted in Fig. \ref{fig:helix_sim}(a).
As a linear spin wave with purely $n_2$  component passes across the right/left wound helix with helicity $\chi=\pm 1$, the spin wave gradually acquires  an out-of-phase (in-phase) $n_3$ component.
Consequently in Fig. \ref{fig:helix_sim}(b), the polarization angles $\varphi$ increase/decreases steadily within the right/left wound circular helix of helicity $\chi=\pm 1$, and level off in  straight wires.
The linear spin wave finally acquires a anticlockwise/clockwise rotation angle of about $0.4 \pi$, in good agreement with theoretical value of $\Delta\varphi=0.392\pi$ calculated from \Eq{eqn:delta_varphi}.
In addition, the spatial profiles for three different spin wave frequencies $f=10, 15, 20~\si{GHz}$ overlap with each other, confirming the frequency-independent feature of spin wave rotation in a circular helix.
By increasing the reduced pitch $b$ from $0.1~\si{\mu m}$ to $0.2~\si{\mu m}$ ($0.3~\si{\mu m}$), the rotation angle also roughly multiply by about $1.9$ ($2.6$) times in Fig.  \ref{fig:helix_sim}(c); and the rotation angles double for a circular helix of $N=2$ coils, as shown in Fig. \ref{fig:helix_sim}(d).
All above behaviors highlight the critical role of the torsion in rotating the polarization direction of a linear spin wave.

\begin{figure}[bt]
	\centering
	%\frame{\includegraphics[width=0.23 \textwidth,trim=140 80 90 100,clip]{figs/curved_AFM.pdf}}
	\includegraphics[width=0.37 \textwidth]{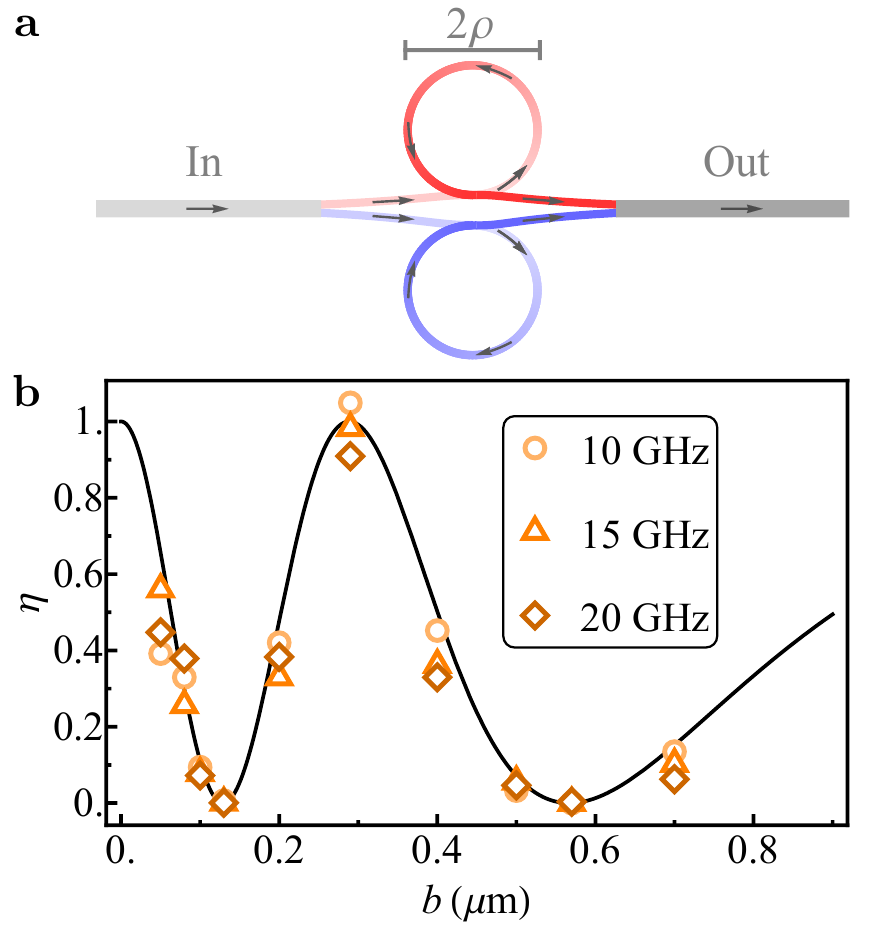}
\caption{{\bf Micromagnetic simulations of the spin wave interference upon two oppositely wound helices.}
(a) Schematics of a spin wave interferometer based on circular helices.
The gray blocks are the input/output ports, the red/blue curves denote two arms upon circular helices of helicity $\chi=\pm 1$,  the darkness encodes the height information, and the arrows mark the flow directions.
(b) The output efficiency $\eta$ as function of reduced pitch $b$.
The dots are extracted from micromagnetic simulations, and the solid line is theoretically calculated from \Eq{eqn:interf}.
In (b), an intensity reduction by a ratio of about $0.34$, caused by signal reflections in imperfect injunctions, is pre-excluded in the estimation of the output efficiency.
\label{fig:two_helix}
}
\end{figure}

\subsection{Spin wave interferometer with two oppositely wound circular helices}
Consider a Mach-Zehnder type spin wave interferometer in Fig. \ref{fig:two_helix}(a), which uses a left/right-handed magnetic helices as its two arms.
Due to symmetry of its two arms, polarized spin wave is split equally into two arms, and merged again in the output port.
According to \Eq{eqn:phase_diff}, the phase shifts of circular spin waves in these two arms are opposite, leading to an overall phase difference of $2|\Delta\phi|$.
Consequently, the interference of spin waves in two arms leads to
\begin{align}
\label{eqn:interf}
       \eta = \cos^2(\Delta\phi)= \cos^2 ( 2\pi \cos\theta ),
\end{align}
where  $\eta$ the output efficiency measuring the ratio of the output intensity relative to the input intensity.
Despite that \Eq{eqn:interf} is originally derived for circular spin waves, it naturally extends to arbitrarily polarized spin wave, for which the intensity is simply the summation of its two circular components \cite{goldstein_Polarized_2017}.

The output efficiency $\eta$ extracted from micromagnetic simulations, as function of reduced pitch $b$ for a fixed helix radius $\rho=0.5~\si{\mu m}$, are overlaid well with theoretical lines explicitly given by $\eta=\cos^2(2\pi b/\sqrt{\rho^2+b^2})$  from \Eq{eqn:interf}.
While the modulation becomes slightly inefficient as $b$ increases due to the accompanying increment of parametric radius $r$, two complete destruction points with $\eta=0$ are established at  $b\approx 0.13~\si{\mu m}$ and $b\approx 0.57~\si{\mu m}$.
Similarly to the spin wave rotation in Fig. \ref{fig:helix_sim}(b), the intensity modulation is the same for three spin wave frequencies $f=10, 15, 20~\si{GHz}$, and we have also additionally checked that the intensity modulations are roughly the same for circularly, linearly and elliptically polarized spin waves.

\section{Discussion and conclusion}
The phase difference $\Delta \phi$ in \Eq{eqn:phase_diff}, which is solely connected to the polar angle $\theta$ but irrelevant to detailed magnetic parameters, is intimately correlated to the Berry phase acquired along the helix \cite{berry_Quantal_1984,dugaev_Berry_2005,xiao_Berry_2010}.
The circular spin wave $|\psi(\bn_0, \sigma)\rangle$ has its  chiral operator $\hat{\bm{\sigma}}$ always parallel to the background magnetization $\bn_0$,  which is governed by
\begin{align}
\label{eqn:berry_sw}
       \bn_0 \cdot \hat{\bm{\sigma}}|\psi(\bn_0(s),\sigma)\rangle = \sigma|\psi(\bn_0(s),\sigma)\rangle,
\end{align}
where the chiral operator  $\hat{\bm{\sigma}}$ is modulated by an effective magnetic field in $\bn_0$.
After travelling along one coil of the circular helix, the staggered magnetization $\bn_0$ restores its original direction, while the spin wave $|\psi(\bn_0(s),\sigma)$ acquires an additional phase factor $\exp(-i\sigma \Omega)$, where $\Omega$ is the solid angle subtended by the static magnetization $\bn_0$ in the Bloch sphere \cite{berry_Quantal_1984,chiao_Manifestations_1986}.
The total solid angle for the cone in Fig. \ref{fig:sch_curved_afm}(b) is $\Omega=2\pi \chi (1-\cos\theta)$, which differs  from \Eq{eqn:phase_diff} only by a trivial value of $2\pi$.
In this sense, the torsion $\tau$ in the curvilinear wire acts as a local Berry curvature as spin wave propagates along a curved wire.

In conclusion,  we demonstrate that the curvilinear effects of magnetic wires provide extrinsic yet universal means to manipulate the polarized spin wave.
Specifically, spin wave rotator and interferometer upon magnetic helices are proposed, due to the additional phase caused by the interplay between the wire winding helicity and the spin wave chirality.
Based on close relations between curvilinearity and magnetism, mechanically controlling magnetic functionalities  is  envisioned.

\begin{acknowledgements}
This work is supported by National Natural Science Foundation of China (Grant No. 11904260) and Natural Science Foundation of Tianjin (Grant No. 20JCQNJC02020).
\end{acknowledgements}
\appendix

\end{document}